\documentclass{edm_article}

\usepackage{hyperref}

\usepackage{appendix}

\begin{document}

\title{SimGrade: Using Code Similarity Measures for More Accurate Human Grading}
%

\numberofauthors{1} 
\author{
\alignauthor
Sonja Johnson-Yu, Nicholas Bowman, Mehran Sahami, and Chris Piech\\
        \affaddr{Stanford University}\\
        \email{\{sonja, nbowman, sahami, piech\}@cs.stanford.edu}
}

\maketitle


\begin{abstract}
While the use of programming problems on exams is a common form of summative assessment in CS courses, grading such exam problems can be a difficult and inconsistent process. Through an analysis of historical grading patterns we show that inaccurate and inconsistent grading of free-response programming problems is widespread in CS1 courses. These inconsistencies necessitate the development of methods to ensure more fairer and more accurate grading. In subsequent analysis of this historical exam data we demonstrate that graders are able to more accurately assign a score to a student submission when they have previously seen another submission similar to it. As a result, we hypothesize that we can improve exam grading accuracy by ensuring that each submission that a grader sees is similar to at least one submission they have previously seen. We propose several algorithms for (1) assigning student submissions to graders, and (2) ordering submissions to maximize the probability that a grader has previously seen a similar solution, leveraging distributed representations of student code in order to measure similarity between submissions. Finally, we demonstrate in simulation that these algorithms achieve higher grading accuracy than the current standard random assignment process used for grading.
\end{abstract}

%


\section{Introduction}
Free-response coding questions are a common component of many exams and assessments in programming courses. These questions are popular because they give students the opportunity to show their understanding of course material and demonstrate their coding and problem-solving skills \cite{doi:10.1111/j.1745-3984.1994.tb00437.x}. However, the flexible nature of these problems introduces unique challenges when it comes to grading student responses, which are compounded in situations where the scale of the course necessitates a team of graders working together (``group grading''). The difficulty of consistent application of grading criteria by a group of graders stems from the incredible diversity of student submissions that are generated for free-response coding questions. In particular, it has been previously shown that the space of different student solutions to free-response programming problems follows a long-tailed Zipf distribution \cite{wu2018zero}. For this reason, it is challenging to develop automated systems for grading and providing feedback and thus human grading remains the gold standard for grading such free-response problems. However, even a team of human graders with extensive experience can struggle to consistently and accurately apply a single, unified criteria when grading. This is problematic as it can result in negative impacts on students in the form of incorrectly assigned grades and inaccurate feedback. Our goal in this paper to explore the frontier of techniques improving the process and outcomes of the exam grading experience.

Our main insight in developing improved approaches for grading is that \textit{it is easier for graders to grade in a consistent manner if they are able to grade similar submissions one after another}. First, we examine historical data to provide concrete evidence of a relationship between grader accuracy and the similarity of previously graded submissions to the current submissions a grader is grading. Then, we propose algorithms that group and order similar submissions in different ways to minimize grader error. Finally, we show that these algorithms perform better than current baseline methods for grading. This work's primary contributions are:
\begin{enumerate}
\item Reporting of grader errors in a CS1 course
\item Using historical data to demonstrate the potential benefits of similarity-based grading 
\item Three algorithms for grading using code similarity
\end{enumerate}

\subsection{Related Work}

\textbf{Autograding}
One commonly used approach to scale grading is the use of autograders \cite{10.1145/1163405.1163407}. While useful for comparing program output for correctness or matching short snippets of code, autograders are more problematic for free-response questions in exam settings. In such contexts, the subtlety of understanding that human graders provide is often essential to providing appropriate feedback to students and properly assessing the (partial) correctness of their solutions. While promising, fully autonomous AI solutions are not ready for grading CS1 midterms \cite{piech2015learning, nguyen2014codewebs, wu2018zero, parihar2017automatic} especially for contexts with only hundreds of available student submissions \cite{wang2017data}.

\textbf{Grading by Similarity} The idea of grouping and organizing student submissions in order to improve grading outcomes has been previously proposed for a variety of problem types. Merceron and Yacef \cite{clusterlogic} use vectors that encode students' mistakes in order to group together students who make similar mistakes when working on formal proofs in propositional logic. Gradescope, designed by Singh et al. \cite{Singh:2017:GFF:3051457.3051466} offers functionality for grading similar solutions, which is currently most effective on multiple-choice-type questions. This approach has also been applied to short answer questions, as explored by Basu et al. \cite{basu2013powergrading}, as well as math problems, as demonstrated by Mathematical Language Processing \cite{mlp}. In this paper, we identify "similar" student responses on free-response programming questions to improve grading quality.

\textbf{Code Similarity} In order to define similarity metrics for student code submissions, we apply techniques for generating numerical embeddings for student programs. Henkel et al.what  \cite{DBLP:journals/corr/abs-1803-06686} created abstracted symbolic traces, a higher-level, light-syntax summary of the programs, and embedded them using the GloVe algorithm \cite{pennington2014glove}. Alon et al. \cite{code2vec} pioneered code2vec, an attention-based embedding model specifically used to represent code. Recently, further advances have been made to improve code embeddings by training contextual AI models on large
datasets from Github \cite{kanade2019learning}. For this application, we favor simpler unsupervised embedding strategies that do not require human-generated labels by adapting the popular NLP technique Word2vec \cite{NIPS2013_5021}, in which ``word'' representations are derived from surrounding context.

\subsection{Dataset}

Our analysis focuses on the student submissions and grader logs from four exams for an introductory programming (CS1) course taught in Python. The breakdown of summary statistics across the four exams is presented in Table \ref{data-table}. As a note, a ``submission'' is defined as one student's written answer to one free-response  problem – thus, the total number of submissions for a given exam is roughly the number of students times the number of coding problems on the exam. In total, we analyze 11,171 student submissions across 1,490 students. Additionally, we have grading logs for every student submission, which consists of information about the grader, the criteria items applied, the final score, and the amount of time that the grader spent on the submission. 199 graders contributed to grading these four exams. As discussed below, the same student submission is sometimes graded by more than one grader for validation purposes.  Thus, our dataset contains 14,597 individual grading log entries.

Our grading data comes from a grading software system that randomly distributes student submissions to graders. Among the standard student submissions for grading, this software also inserts ``validation'' submissions that have already been graded by senior teaching assistants. Every grader assigned to a specific problem will grade all ``validation'' submissions for that problem. The presence of these special submissions creates opportunities for assessing grader performance, both relative to their peers and relative to ``expert" performance.

\begin{table}
\begin{tabular}{ cccc } 
 Exam \# & \# Students & \# Submissions & \# Graders \\ \hline
  1 & 533   & 3,731  & 53 \\ 
  2 & 259	& 1,813  &	52\\ 
  3 & 247   & 2,470	    &	51\\ 
  4 & 451   & 3,157      &	43 \\ \hline
  Total & 1,490 & 11,171  & 199 \\
 \hline
\end{tabular}
\caption{Exam Grading Dataset Summary Statistics}
\label{data-table}
\end{table}

\section{Natural Grading Error} 
\begin{figure}[h]
\begin{center}
      \includegraphics[width=0.8\linewidth]{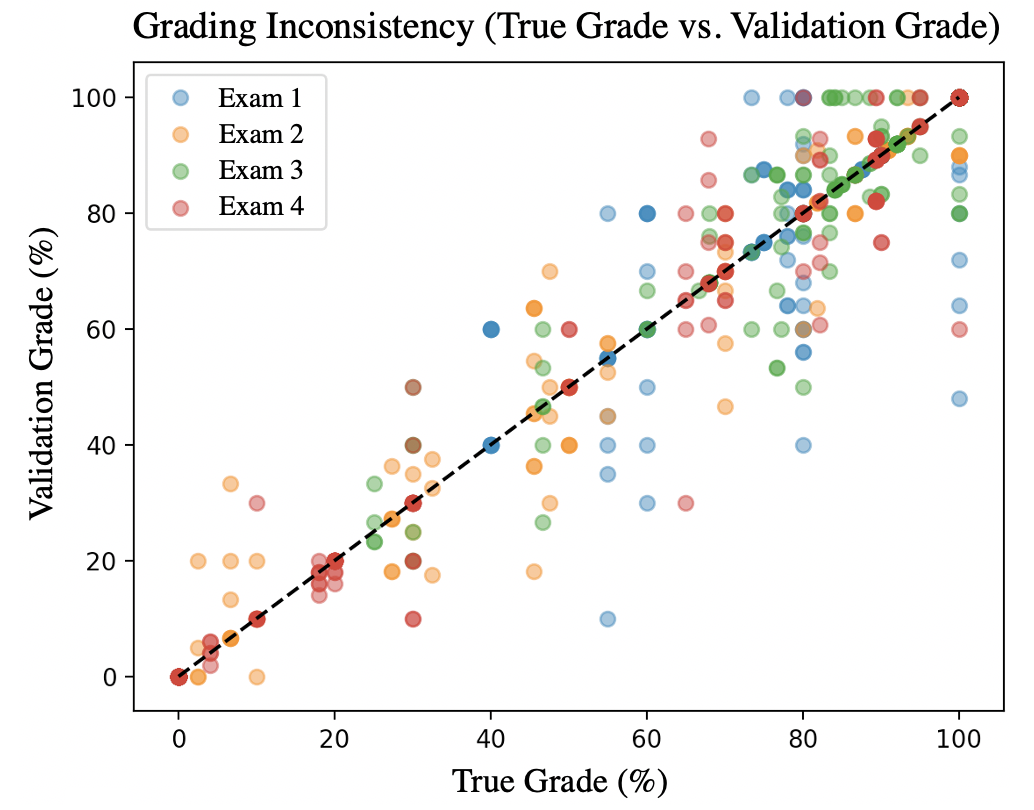}
  \caption{True grade assigned by expert vs. validation grade assigned by human grader} 
  \label{inconsistent-grading}
\end{center}

\end{figure}

 While anecdotal experience of grading inconsistency is a common trend in our experience as educators, our first focus is to quantify the inconsistencies present in historical grading sessions in a rigorous manner. In particular, our analysis focuses on the aforementioned ``validation'' submissions that were specially handled by the grading software and assigned to every grader working on a specific problem. As a result, we had a subset of the grading logs for which we knew both the true grade (as defined by an expert) and the ``validation'' grade assigned by each grader. Plotting these values against one another is shown in Figure \ref{inconsistent-grading}, which reveals troubling inconsistencies in the grades assigned by graders.  With an RMSE of 7.5 (i.e., average error of 7.5 percentage points per problem), we see that grading error is significant, nearly on the order of what would translate to a full letter grade. Linear regression on this plot yields an R-squared coefficient of 0.947 indicating that while the error may be high, the direction of errors is generally unbiased.  In other words, there is not systematic over/under-grading.  Rather, the grading errors tend to be randomly distributed around the true grade. Thus, the rest of this paper focuses on methods for decreasing this demonstrated inconsistency (absolute error) in human grading.

\begin{figure*}[h!]
    \centering
    \includegraphics[width=1.0\textwidth]{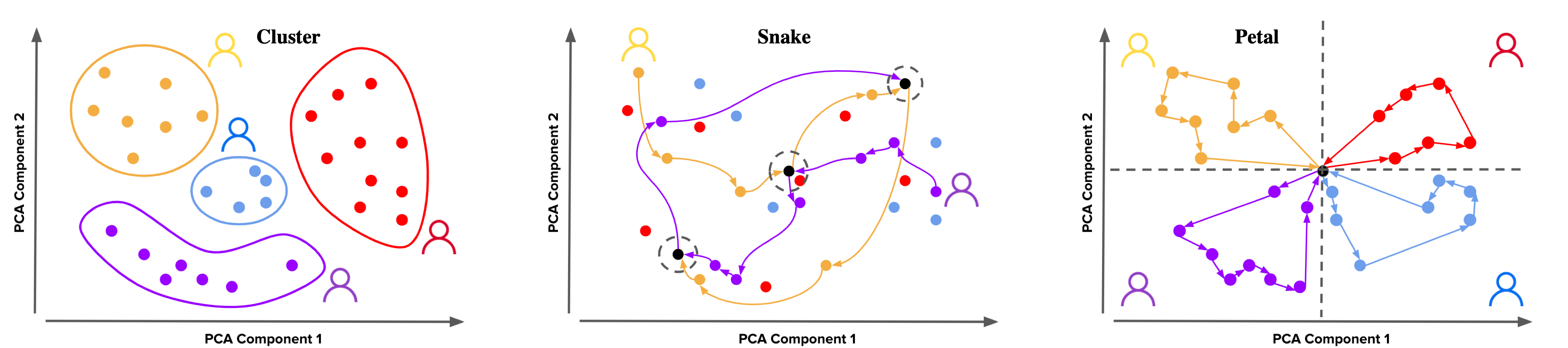}
    \caption{Submission assignment via three algorithms: Cluster, Snake, Petal}
    \label{fig:algos}
\end{figure*}

\section{Methods}

In this section, we will first outline methods for answering key questions about the problem of improving human grading using similarity scores. Then, we will present three novel algorithms for improving human grading.

\subsection{Can code similarity be accurately captured?}  

We generate program embeddings for all student submissions in our corpus. Word embeddings are an established method of encoding semantics in human language \cite{NIPS2013_5021, pennington2014glove, DBLP:journals/corr/abs-1810-04805, DBLP:journals/corr/abs-1810-04805}, and these same techniques applied to code accomplish similar results. Algorithms for generating embeddings are constantly evolving and improving; to avoid over-optimization at the embedding generation stage, we chose to employ the simple baseline Word2Vec algorithm. We then demonstrated that our embeddings are semantically significant using zero-shot rubric sampling \cite{wu2018zero}. For details, see the Appendix\footnotemark.

\footnotetext{
\url{https://compedu.stanford.edu/papers/appendices/SimGradeAppendix.pdf}}

\subsection{Does similarity influence grader accuracy?}

We hypothesize that graders score submissions more accurately when they have recently seen a submission similar to the current submission. To test this hypothesis, we analyze grading data for four exams.  First, for each grader, we generate a ``percentage grading error,'' which is an average of their absolute percent deviation from the correct answer on all validation submissions that they graded. Then, for each of the validation submissions that a grader evaluated, we sort their personal grading logs by time and look at the window of three submissions leading up to each validation submission they graded. To quantify similarity of the validation submission to recently graded submissions, we take the maximum of the cosine similarity between the current validation submission and the three previous submissions. We plot the maximum similarity between a validation submission and the previous submissions against a grader's percentage grading error in order to identify the relationship between a grader's history and accuracy. Then we can infer a formula that approximates the relationship between previous submission similarity and percentage grading error.

\subsection{Algorithms to assist human grading}

We compare four algorithms for assigning submissions to graders: (1) Random, in which submissions are randomly assigned to graders, with five ``validation'' submissions interspersed for assessing grader bias. This is the status quo and serves as the baseline. (2) Cluster, in which each grader is assigned to a cluster of highly similar submissions. (3) Snake, in which each grader is randomly assigned a set of submissions and is shown the submissions greedily by nearest neighbor. (4) Petal, in which the dataset is divided into ``petals'' and all graders begin in the same place. Figure \ref{fig:algos} provides a visualization of (2), (3), and (4). Detailed explanations of the algorithms are in the Appendix\footnotemark[\value{footnote}].

\begin{figure}[h]
\begin{center}
    \includegraphics[width=0.7\linewidth]{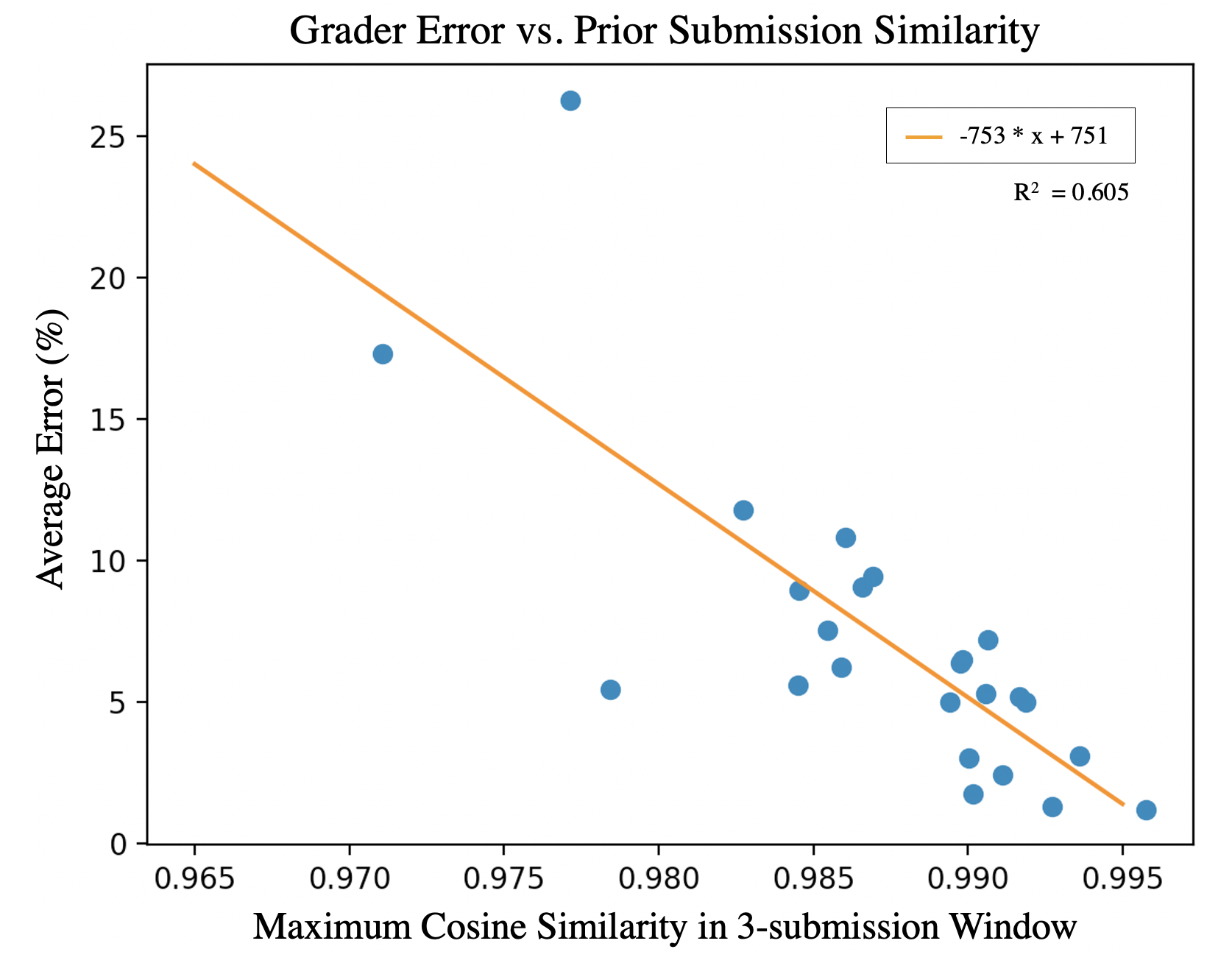}
  \caption{Relationship between grader accuracy and similarity in 3-submission window prior to validation submission}
  \label{similarity-accuracy}
\end{center}

\end{figure}

\begin{figure*}[h]
    \centering
    \includegraphics[width=1\textwidth]{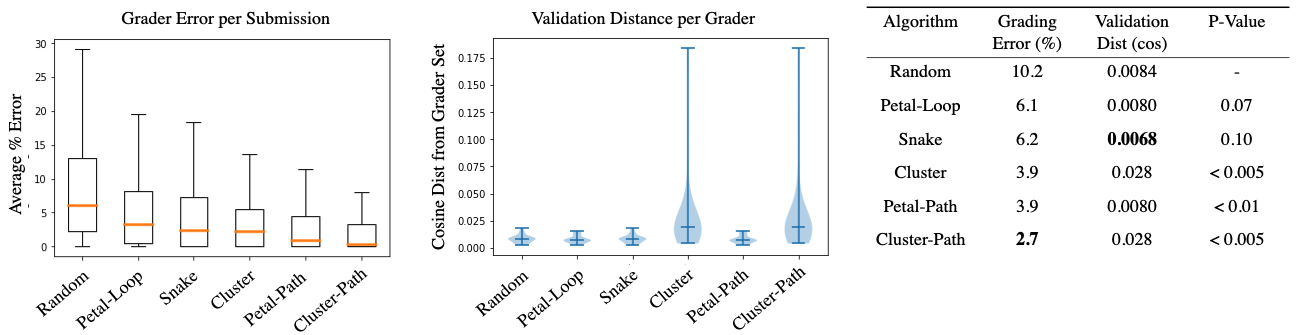}
    
    \caption{Left: Average per-submission grading error for each algorithm, Center: Distance of validation submissions from normally assigned submissions, Right: Summary performance statistics, including comparison to random baseline.}
    \label{fig:sim-results}
\end{figure*}

\subsection{Algorithm evaluation}

To evaluate the performance of the different algorithms, we simulate grading for a 444-person six-problem exam and ten graders, using real student programs from an actual exam. Details about the selection of validation submissions are in the Appendix\footnotemark[\value{footnote}]. When running the simulation, we infer percentage grading error by examining the similarity of the previous three submissions to the current submission. While we emphasize grader error as the most important metric for assessing an algorithm, a secondary consideration is how naturally validation submissions integrate with the rest of a grader's assigned submissions. Ideally, a validation submission is not ``out-of-distribution'' with respect to the other submissions that a grader is assigned. Otherwise, a grader will be able to tell when they are being evaluated for grading accuracy. To assess how ``out-of-distribution'' the validation submissions are, we examine how dissimilar the validation submissions are from the non-validation submissions assigned to a grader. Specifically, for each validation submission, we measure the distance between the validation submission and the nearest non-validation submission assigned to that grader. We average over the five validation submissions in order to get the mean minimum distance from validation to non-validation for a grader, which will be higher if one of the validation submissions is out-of-distribution.

\section{Experimental Results}

\subsection{Similarity scores are meaningful}

Embeddings are semantically significant because similarity between embeddings corresponds to similarity between submission feedback labels, as described in the Appendix\footnotemark[\value{footnote}].

\subsection{Similarity influences grading}
\label{section-similarity-accuracy}

Graders score assignments more accurately when they have recently seen a submission similar to the current submission they are grading. From our analysis of historical data, we find that when there is a high similarity between the current submission and at least one of the previous three submissions, the percentage grading error is low. Conversely, when the similarity between previous submissions is low, the percentage grading error is high. We find a linear relationship between the maximum similarity of the previous three submissions and the percentage grading error as shown in Fig. \ref{similarity-accuracy}, with $R^2 = 0.605$. Given that the grading process involves the numerous uncertainties that come along with human involvement, we believe this correlation coefficient shows a statistically significant relationship between historical similarity and grader accuracy. While the linear relationship between historical submission similarity and percentage grading error is a simplifying assumption, it is the best assumption we can make given evidence provided in Fig. \ref{similarity-accuracy}.

\subsection{Improved accuracy by algorithm}

We compare six algorithms for assigning submissions to graders and selecting an order in which a grader will view a submission in Figure \ref{fig:sim-results}. We apply the equation of the linear relationship shown in Figure \ref{similarity-accuracy} to the similarity of submissions as ordered for evaluation by different algorithms in our experiments. This equation allows us to predict grader accuracy when using the orderings provided by different algorithms. We find that implementing a path ordering on a clustered assignment of graders to submissions yields the lowest mean error of 2.7\% (bold-ed in Fig. \ref{fig:sim-results}), while the other algorithms all show an improvement over the baseline 10.2\% grading error. We utilize bootstrapping \cite{Efron_1979} over 100,000 trials in order to get the p-values that indicate the significance of the difference in means between the baseline algorithm and the other algorithms (see table in Fig. \ref{fig:sim-results}).

\subsection{Validation viability by algorithm}

When comparing the cluster, snake, and petal algorithms, we observe that the cluster-based algorithms are most likely to have validation submissions that are ``out-of-distribution,'' with a mean validation distance of 0.0277. All other algorithms have substantially lower mean minimum distances.

\section{Discussion}

Overall, we saw that all of our novel proposed algorithms for assignment of submissions to graders provided improvements over the random baseline in simulation. In general, we saw that path-based algorithms (petal-path and cluster-path) had lower grading error than their non-path counterparts because they are designed to optimize for maximum similarity between consecutive submissions that a grader grades. In particular, the cluster-path algorithm yielded the lowest grader error in simulation due to its strong tendency to assign very similar submissions to graders. On the other hand, the snake algorithm provided the most optimal average distance to validation submissions, which may be important for a smooth experience for a real-life grader. Finally, we saw that the petal algorithm offered a balanced trade-off between these two extremes – while not optimal in either metric, it can be a good choice when both metrics (grading error and validation submission distance) are equally important for designing a grading experience. For a more in-depth discussion of our observed results, see the Appendix\footnotemark[\value{footnote}].

\section{Conclusion}
Through analysis of historical exams, we demonstrated that there is inconsistency between true scores and grader-assigned scores. In doing so, we introduce a new task and associated measure, \emph{grading correctness}. Moreover, we found experimental support for our hypothesis that graders are able to assign scores to exam problems more accurately when they have previously seen similar submissions.  In turn, we proposed the use of code embeddings to capture semantic information about the structure and output of programs and identify similarity between submissions. Using similarity of code embeddings in conjunction with historical grading data, we demonstrate in simulation that graders are indeed able to score submissions more accurately when they have previously seen another submission similar to it. We propose and compare several algorithms for this task, showing that it is possible to achieve a significant increase in grading accuracy over simple random assignment of submissions. Future extensions of this work include (i) improvements on code embeddings and (ii) deployment of the grading algorithms in an operational system to allow more direct experimental comparison of grading accuracy. The use of such algorithms show promise for improving accuracy, and in turn fairness, in evaluations of student performance.

%
\bibliographystyle{abbrv}
\bibliography{sigproc}  
%
%

\newpage
\appendix

\section{Methods}

\subsection{How accurate are similarity scores?}

We generate our embeddings with Word2vec, an unsupervised embedding technique used commonly in NLP. During the research process, we experimented with other types of embeddings, including simple ``count'' and ``presence'' vectors that simply tabulated the presence and amount of different tokens in the code (i.e., ``bag-of-words'' representations) and more complex contextualized embeddings, such as those generated by BERT. Out of these different embeddings, we chose to proceed with the Word2Vec embeddings due to their balanced trade-off of expressiveness and ability to be trained on a reasonable amount of data. Word2vec learns low-dimensional vector representations of words by using the context of each word to create its representation. The code embedding for a program is an average of the embeddings for its component tokens.


We pre-process the Python student programs by stripping out comments, inserting semicolons and braces to indicate line breaks and control flow statements, masking out long strings, and tokenizing the programs. Tokens that appear fewer than five times across all student submissions for a problem are removed, since the representations for rare tokens tend to be less semantically rich.

We evaluate the semantic quality of our code embeddings by looking at pairs of embeddings, examining the relationship between embedding similarity and label similarity. Here, label similarity refers to the Jaccard similarity between two submissions' grading criteria feedback. The Jaccard similarity between two sets is defined as $J(A,B) = \frac{|A \cap B|}{|A \cup B|}$. For example, if one submission received the feedback labels [`off-by-one error', `doesn't return a value'] and another received the feedback labels [`off-by-one error', `doesn't return a value', `incorrect loop condition'], the Jaccard similarity would be $\frac{2}{3}$, since the two submissions share two labels out of three unique labels in total.

We measure similarity between program embeddings using cosine similarity, a common metric for assessing word embedding similarity, defined as $\text{similarity}(\vec{a}, \vec{b}) = \frac{\vec{a} \cdot \vec{b}}{||\vec{a}|| ||\vec{b}||}$. The cosine similarities for our embeddings range from 0.80 to 1.00 because all of the programs share a certain degree of structural similarity (such as the function header, which was provided) and because we generate program embeddings by averaging over the embeddings for the various tokens.

We used zero-shot rubric sampling to explore the relationship between measured similarity and true similarity. Since there is no single ``ground truth'' regarding how similar two real submissions truly are, we turn towards analyzing embedding performance on simulated data, whose true similarity can be precisely known. Since grader feedback is often too coarse to decipher structural information about a program, it makes sense to simulate the relationship between student decisions and the code they produce. Because our student exam data is graded with coarse, qualitative feedback labels (such as ``Minor error'' and ``Major error'']) and is subject to grading error, we synthesize labeled data by writing a grammar that can generate over 120,000 unique solutions. These synthetic submissions are labeled with descriptions of the output of the program as well as significant structural features (e.g. the use of a for loop vs. none). Since the labels contain important semantic information, label similarity is a useful heuristic for semantic similarity.

\begin{figure}[]
\begin{center}

  \includegraphics[width=0.7\linewidth]{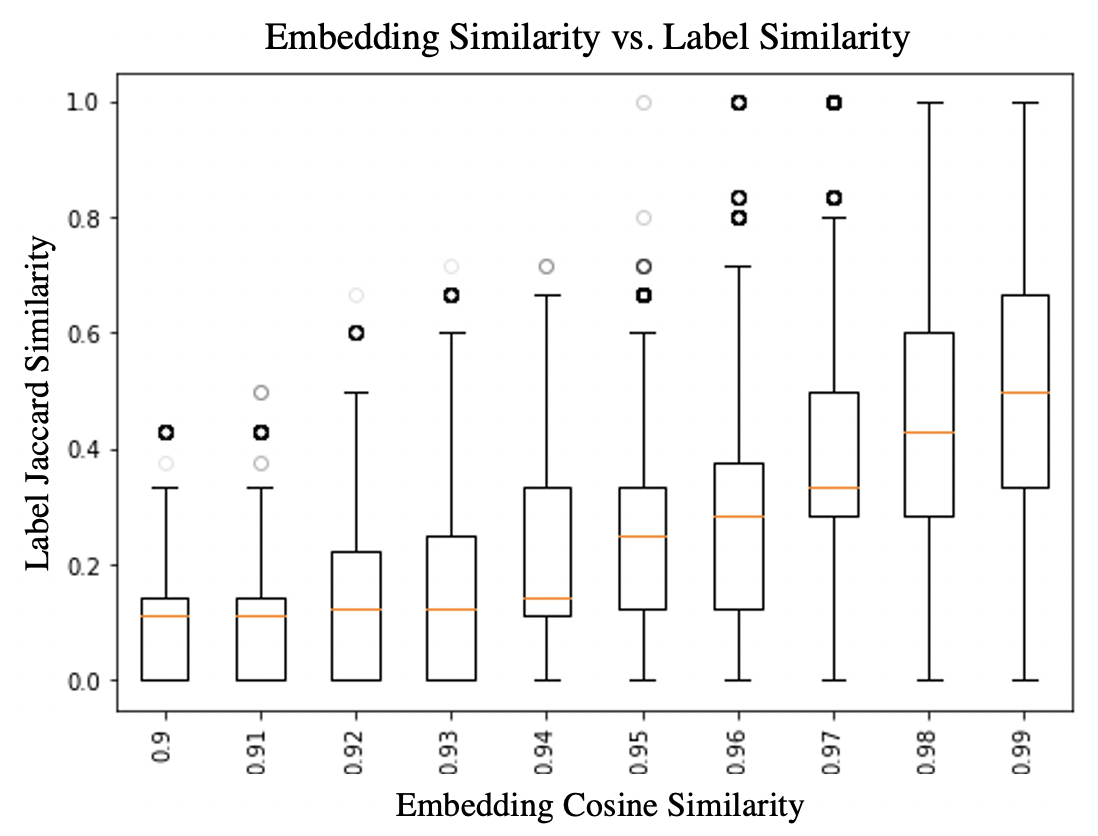}
  \caption{Notable relationship between calculated similarity of embeddings and true similarity of labels }
  \label{embedding-semantics}
  \end{center}
\end{figure}

\subsection{Algorithms}

\subsubsection{Random (Baseline)}

In a baseline grading session, submissions are randomly assigned to a grader, and each of the five validation submissions are randomly interleaved among a grader's non-validation submissions. Graders view submissions in a random order. This approach is most common in practice today.

\subsubsection{Cluster}

In a clustered grading session, each grader is assigned to a cluster of highly similar submissions, as shown in the left-hand plot of Figure \ref{fig:overview}. We generate these clusters using K-means clustering \cite{1056489}, where $k$ equals the number of graders. K-means clustering assigns items to clusters, aiming to maximize the similarity of the items within a cluster. Since K-means clustering uses cosine similarity to group embeddings, our resulting clusters have semantically similar programs, so that one grader is responsible for grading many similar programs. In addition to the submissions within a cluster, the grader is also responsible for grading five validation submissions. We implement two methods for ordering submissions within a cluster: (1) random, in which the submissions are displayed in a random order, and (2) path, in which a random submission is chosen as a starting point, and the following submissions are ordered by continually selecting the nearest neighbor, measured by cosine similarity. 

\subsubsection{Snake}

Like the baseline random algorithm, the ``snake'' algorithm begins by randomly assigns a set of submissions to each grader, along with the five validation submissions. Unlike the baseline algorithm, however, the snake algorithm picks a random starting point within the set assigned to a grader and then traverses through the assigned submissions greedily by ordering the submissions according to nearest neighbor. This creates a path throughout the solution space that attempts to maximize similarity between subsequent submissions.

\subsubsection{Petal}

The ``petal'' algorithm stems from the insight that there are some situations when it is valuable for a grader to end in the same place where they begin. One benefit is that if all graders begin and end in the same place, and one grader finishes grading more quickly than the others, she can begin working on another grader's ``petal'' without needing to jump to a vastly different set of submissions, which could introduce greater grader error. The ``petal'' algorithm takes a clustering-like approach to dividing the data. Instead of clustering, however, it divides the dataset into $k$ petals, where $k$ equals the number of graders. First, the embeddings are reduced to two dimensions via principle component analysis (PCA). and standardized to have mean 0 and variance 1. Next, they are split into $k$  ``petals'' corresponding to equally-sized angles. Each petal is a cycle which shares a common node with other cycles. This ensures that a grader's submissions are semantically similar to one another and not too semantically distant from another grader's submissions.

We take two approaches for ordering a grader's submissions within the petal dataset. Finding a shortest path circuit through $n$ points, starting and ending in the same place, is an NP-hard problem known as the traveling salesman problem. We use a stochastic heuristic algorithm for generating a solution, implementing Markov chain Monte Carlo (MCMC) to create an ordering. Additionally, we implement a second approach that generates a path as opposed to a loop, greedily ordering by nearest neighbor.

\subsection{Algorithm Evaluation}
 For each exam problem, we randomly select five validation submissions out of the 444 total submissions. The 439 non-validation submissions are divided into 10 subsets (each for one grader), the 5 validation submissions are added, and each subset of submissions is ordered according to the given algorithm.

\section{Discussion}

All proposed algorithms improve upon the random baseline's grading error, though each has its own advantages.

The cluster-path algorithm yields the lowest percentage grading error because K-means clustering is designed to make submission subsets of high similarity, and because the path is created by greedily choosing the most similar submissions, thereby minimizing error. In general, path-based algorithms have lower grading error than their non-path counterparts because they are designed to optimize for maximum similarity between consecutive submissions that a grader grades. Under our assumed model of grading error, in which higher submission similarity corresponds to lower grading error, these approaches will minimize overall grader error.

The snake algorithm yields the lowest mean distances from validation to non-validation submissions, rendering it the best choice for a smooth grader experience. Because validation submissions are randomly sampled and the snake algorithm traverses the entire submission space, its validation submissions are not far from non-validation submissions. 

The petal algorithm, while not optimal in grading error or validation distance, offers a compromise between the benefits and drawbacks of the other two algorithms – it is best used when both metrics are equally important. This algorithm traverses a wider space of submissions than cluster does, but it groups similar solutions, unlike snake.


Finally, we consider the real-life situations in which these algorithms would be applied. In particular, there are certain aspects of live grading sessions that complicate the assignment of submissions to graders. One is the possibility that certain graders might finish before others and might need to be reassigned elsewhere in order to maintain optimal grading throughput. In these cases, the ``finished'' grader would generally start working on submissions assigned to another, slower grader. If the submissions were initially assigned by the cluster algorithm, the ``finished'' grader could end up switching to a very different cluster of submissions composed of substantially different programs, which would make them likely to have substantial grading error at first. This impact would be mitigated if submissions were assigned by the snake algorithm, since a single grader's submissions are randomly distributed over the submission space, which means that another grader's submissions are likely close by. The petal algorithm also mitigates ``switching'' costs by having all graders begin and end around the most ``central'' submission, so that a finished grader can smoothly continue onto another grader's petal.

In sum, while the cluster algorithm produces the lowest grading error, the other algorithms may be more useful in practice when considering the real-world impacts of choosing good validation submissions and dynamically balancing grader load in an optimal manner. In particular, the snake algorithm yields the the best opportunity to seamlessly integrate validation submissions (lowest distance to validation submissions). Finally, the petal algorithm offers a middle-ground trade-off between the cluster and snake algorithms.

\section{Experimental Results}

\subsection{Similarity scores are meaningful}

We compared the embedding similarities with label similarities for 499,019 unique pairs of programs generated by our grammar. In Figure \ref{embedding-semantics}, we identify a linear relationship between the cosine similarity of two code embeddings and the Jaccard similarity between their labels, indicating that the use of code similarity measures to select and order problems for human graders to evaluate show promise for improving the accuracy of grading.  Even when generated with a simple algorithm such as Word2vec, these embeddings carry valuable semantic information in determining the structure and correctness of programs.  Moreover, as methods for generating code embeddings continue to improve in the future, it will likely lead to even better human grading consistency when used such embeddings in conjunction with the approach to grading outlined in this work.

\balancecolumns
\end{document}